\documentclass[useAMS,usenatbib]{mn2e}
\usepackage{amssymb}
\usepackage{amsmath}
\usepackage{times}
\usepackage{graphicx}

\newcommand{\bv}{\mbox{\bf v}}
\newcommand{\bs}{\mbox{\bf s}}
\newcommand{\bl}{\mbox{\bf l}}
\newcommand{\lya}{Ly$\alpha$\,}

\newcommand{\Lya}{\mbox{Ly$\alpha$}}

\newcommand{\nhi}{\mbox{$N_{\rm HI}$}}
\newcommand{\msun}{\mbox{$M_\odot$}}

\newcommand{\kms}{\mbox{km s$^{-1}$}}
\newcommand{\hmpc}{\mbox{Mpc $h^{-1}$}}
\newcommand{\cm}{cm$^{-2}$}

\newcommand{\h}{$h^{-1}_{100}$}

\title[The relation between Lyman--$\alpha$ absorbers and gas--rich galaxies in the local universe]
{The relation between Lyman--$\alpha$ absorbers and gas--rich galaxies in the local universe}

\author[Marco Pierleoni, Enzo Branchini \& Matteo Viel]
{Marco Pierleoni$^{1}$, Enzo Branchini$^{2}$ \& Matteo  Viel$^{3,4}$\\
$^1$ Max-Planck-Institut f\"ur Astrophysik, Karl-Scharzschild Strasse 1, D-85740 Garching, Germany \\
$^2$ Dipartimento di Fisica, Universita\`a degli Studi ``Roma Tre'', via della Vasca Navale 84, I-00148 Roma, Italy\\
$^3$ INAF - Osservatorio Astronomico di Trieste, Via G.B. Tiepolo 11,
I-34131 Trieste, Italy\\
$^4$ INFN/National Institute for Nuclear Physics, Via Valerio 2,
I-34127 Trieste, Italy\\
}

\begin{document}
\maketitle
\label{firstpage}

\begin{abstract}
We use high-resolution hydrodynamical simulations to investigate the
spatial correlation between weak (\nhi\ $<10^{15}$ \cm)
\lya\ absorbers and gas--rich galaxies in the local universe.  We
confirm that \lya\ absorbers are preferentially expected near
gas--rich galaxies and that the degree of correlation increases with
the column density of the absorber. The real--space galaxy
auto--correlation is stronger than the cross--correlation
(correlation lengths $r_{0,gg}=3.1\pm 0.1 \,\hmpc$ and
$r_{0,ag}=1.4\pm 0.1\, \hmpc$, respectively), in contrast with the
recent results of \citep[RW06]{RW06}, and the auto--correlation of
absorbers is very weak.  These results are robust to the presence of
strong galactic winds in the hydrodynamical simulations. In
redshift--space a further mismatch arises since at small separations
the distortion pattern of the simulated galaxy-absorber
cross-correlation function is different from the one measured by RW06.  
However, when sampling the intergalactic medium along a limited number of
lines--of--sight, as in the real data, uncertainties in the cross
correlation estimates are large enough to account for these
discrepancies. Our analysis suggests that the 
statistical significance of difference between the
cross--correlation and auto--correlation signal in current datasets 
is $\sim1$-$\sigma$ only.
\end{abstract}


\section{Introduction}
\label{sec:intro}
Understanding the interplay between galaxies and the intergalactic
medium (IGM) is a fundamental cosmological problem.  On one side the
IGM acts as reservoir of gas that cools down in the potential wells of
dark matter haloes and forms galaxies and stars. On the other side the
IGM is a sink that records, over a large fraction of the cosmic time,
the crucial thermal and chemo--dynamical processes related to galaxy
formation.  Significant progress has been made in the last few years
thanks to high--resolution spectroscopic data from quasar (QSO)
lines--of--sight and imaging of QSO fields that has been performed by
several groups. The properties of \lya and metal absorption lines in
the high redshift universe have been cross--correlated with those of
the galaxies (e.g. \cite{Adelberger05, Nestor07, Bouche07, Schaye07,
  Churchill07}) to shed light on the physical state of the IGM around
them and possibly on the still poorly understood feedback
mechanisms. Among all the possible elements in various ionization
stages hydrogen is the most abundant and thus has been widely studied
by the scientific community.  The analysis of the statistical
properties of \lya lines and of the transmitted flux shows that the
neutral hydrogen in the high--redshift universe is embedded in the
filamentary cosmic web that traces faithfully, at least on large
scales, the underlying dark matter density field (for a review see
\cite{Meiksin07}). At lower redshifts, the situation is likely to be
more complicated (e.g. \cite{Dave03}): the non--linear evolution of
cosmic structures changes the simple picture above allowing \lya
absorbers to populate a variety of environments from the large scale
structure to galaxy groups and underdense regions
(e.g. \cite{LeBrun96,Penton02,Rosenberg03,Lanzetta96,Bowen02,McLin02,Grogin98,Cote05,Putman06}).
Furthermore, because of the atmospheric absorption of UV-photons, the
low redshift \lya absorbers  can be studied only from space based observatories
(\cite{Weymann98,Tripp02}) on a limited number of lines--of--sight
making the results potentially affected by cosmic variance and/or
small number statistics.  The cross--correlation function between low
redshift galaxies and \lya absorbers is the cleanest statistic for
quantifying the relation between the two populations and has been
investigated recently both observationally and using some
hydrodynamical simulations (\cite{Chen05,RW06,Wilman07}), with
somewhat contradictory findings. RW06 using the HI Parkes All Sky
Survey (HIPASS) data set (\cite{Meyer03,Wong06}) has found a puzzling
result: the galaxy--absorber cross--correlation signal is stronger
than the galaxy auto--correlation on scales 1-10 $h^{-1}$ Mpc.
Earlier studies, based, however, on a limited sample of 16 \lya
lines-of-sight, showed the opposite trend (\cite{Morris06}).  The RW06
result is not well reproduced either observationally or theoretically
by \cite{Wilman07} who relied on a different data set
(\cite{Morris06}) and considered a single hydrodynamical simulation. 
The results of \cite{Chen05} seem to be more consistent with the findings of
\cite{Wilman07}.  However, it is worth stressing that while the RW06
galaxy sample includes low redshift objects the other two have been
obtained from magnitude limited catalogs at  higher redshifts.

In this paper we compute the auto and cross--correlation functions of more
than 6000  \lya\  absorbers 
over $\sim 1000$ independent lines--of--sight
and $\sim 5000$ mock galaxies
extracted from the $z=0$ outputs of three different high-resolution hydrodynamical
simulations of a $\Lambda$CDM universe in order to better investigate the issues above.  

In Section~\ref{sec:simu} we
present the numerical experiments and describe the samples of
simulated galaxies and \lya\ absorbers.  The details of the auto and
cross-correlation analyses are described in Section~\ref{sec:method}.
The correlation analysis of the mock samples of galaxies and absorbers
is performed in real--space (Section~\ref{sec:rspace}) and redshift
space (Section~\ref{sec:zspace}). The results are then summarized and
discussed in Sections ~\ref{sec:dis} and ~\ref{sec:per}. 

\section{Hydrodynamical Simulations and Mock Samples}
\label{sec:simu}

We use a set of three hydrodynamical simulations run with {\small
  {GADGET-2}} and its new fastest version {\small {GADGET-3}}, a
parallel tree Smoothed Particle Hydrodynamics (SPH) code that is based
on the conservative `entropy--formulation' of SPH
\citep{SpringelHernquist02,Springel05}. The simulations cover a
cosmological volume (with periodic boundary conditions) filled with an
equal number of dark matter and gas particles.  Radiative cooling and
heating processes are followed for a primordial mix of hydrogen and
helium following the implementation of \cite{KWH}. We assume a mean
Ultra Violet Background (UVB) produced by quasars and galaxies as
given by \cite{HM96}, with the heating rates multiplied by a factor
$3.3$ in order to better fit observational constraints on the
temperature evolution of the Intergalactic Medium (IGM) at high
redshift. {Multiplying the heating rates by this factor (chosen
  empirically) results in a larger IGM temperature at the mean density
  which cannot be reached by the standard hydrodynamical code but aims
  at mimicking, at least in a phenomenological way, the
  non-equilibrium ionization effects around reionization (see for
  example \cite{Bol07}).  The star formation
criterion for one of the simulations (No Winds -- NW) very simply
converts all gas particles whose temperature falls below $10^5$ K and
whose density contrast is larger than 1000 into (collisionless) star
particles, while for other two simulations with strong galactic winds
(Strong Winds -- SW and Extreme Strong Winds -- ESW) a multiphase star
formation criterion is used.

The implementation of galactic winds is described in
\cite{SpringelHernquist03} but we summarize here the main
features. Basically, the wind mass-loss rate $\dot{M}_W$ is assumed to
be proportional to the star formation rate, and the wind carries a
fixed fraction $\chi$ of the supernova (SN) energy. Gas particles are
stochastically selected and become part of a blowing wind, then they
are decoupled from the hydrodynamics for a given period of time or
till they reach a given overdensity threshold (in units of $\rho_{th}$
which is the overdensity threshold for star formation) in order to
effectively travel to less dense regions. Thus, four parameters fully
specify the wind model: the wind efficiency $\eta$, the wind energy
fraction $\chi$, the wind free travel length $l_w$ and the wind free
travel density factor $\delta_w$. The first two parameters determine
the wind velocity $v_w$ through the following equations:
\begin{equation}
\dot{M}_w= \eta \dot{M}_{\star},
\end{equation}
and
\begin{equation}
\frac{1}{2} \dot{M}_w  v_w^2 = \chi \epsilon_{\rm SN} \dot{M}_{\star},
\end{equation}
from which one can compute the maximum allowed time of the decoupling
$t_{dec} = l_w / v_w$.  The parameter $l_w$ has been introduced in
order to prevent a gas particle from getting trapped into the
potential well of the virialized halo and in order to effectively
escape from the Inter Stellar Medium (ISM), reach the low density IGM
and pollute it with metals.  We used similar values to those that have
been adopted by recent studies (e.g. \cite{Nagamine07}) that found
that the outcome of the simulation is relatively insensitive to the
choice of this parameter.  We note that this wind implementation is
different from the momentum--driven implementation of \cite{oppe06},
which seems to better fit statistics of CIV absorption in the
high--redshift universe.

\begin{table}
\begin{center}
\begin{tabular}{lccccc}
\hline 
\multicolumn{6}{|c|}{Simulations} \\ \hline
\small{Run} & \small{$v_W$ (km/s)} & \small{$\chi$} & \small {$\eta$} & \small{$\delta_w$} & \small {$l_W$ (kpc)}\\
\hline
NW   & --    & --  & --  & -- & -- \\ \hline
SW   & 484  & 1  & 2 & 0.1 & 20 \\ \hline
ESW  & 484  &  2  & 4 & 0.025 & 60 \\ \hline
\hline
\label{table:sim}
\end{tabular}

\caption{Main parameters of the simulations. NW (No Winds) uses the quick 
  option for the star formation criterion that converts all the gas
  particles below $10^5$ K and above $\delta=1000$ into stars.  SW
  (Strong Winds) and ESW (Extremely Strong Winds) models use the default
  multiphase star formation criterion.  The density $\rho_w =
  \delta_w\, \rho_{th}$ denotes the threshold density for the
  decoupling of the hydrodynamic force, and $l_w$ indicates the wind 
  free travel length.}   
\end{center}
\end{table}

Throughout $h$ indicates the Hubble constant at the present epoch,
$H_0$ in units of $100$ \kms\ Mpc$^{-1}$.
The cosmological model corresponds to a `fiducial' $\Lambda$CDM
Universe with $\Omega_{\rm m
}=0.26,\ \Omega_{\Lambda}=0.74,\ \Omega_{\rm b}=0.0463$, $n_s=0.95$,
 $H_0 = 72$ km s$^{-1}$ Mpc$^{-1}$ and $\sigma_8=0.85$ (the B2
series of \cite{Viel04}). These parameters provide a good fit to the
statistical properties of transmitted \lya\ flux at $z>2$.  We use
$2\times 400^3$ dark matter and gas particles in a volume of size
$60\ h^{-1}$\,Mpc box and the simulations are evolved down to
$z=0$. The gravitational softening is set to $5\,h^{-1}$\,kpc in
comoving units for all the particles. The mass per gas particle is
about $4.3\times 10^7 M_{\odot}$ which is a factor $\sim 5$ better than
that of \cite{Wilman07}.

These three simulations offer us the opportunity to investigate the
galaxy--IGM interplay at $z=0$ taking into account the role of
different amount of feedback in the form of galactic winds and the
role of two different criteria of star formation. Note that similar
investigations using the same hydrodynamical code and focussing on the
properties of neutral hydrogen around Damped \lya\ systems have been
performed by \cite{Nagamine07}.  In Figure~\ref{fig:simul} 
we present a
qualitative view of
 the neutral hydrogen overdensity in a slice of
thickness 6 comoving $\hmpc$ for the ESW run.
We note a clear tendency for neutral hydrogen to avoid
hot environments, where the neutral fraction is lower.
The HI distribution in the  NW and SW simulations 
it is almost identical on the scale of the plot, and therefore are
not shown here.
Differences can only be spotted on scales smaller than 0.5
comoving \hmpc in which  compact knots of 
neutral hydrogen are seen in the  ESW that are not present in the NW
simulation, since the simplified star formation criterion of this
latter converts cold gas into collisionless stars. 
We will address the differences between the simulations in a
quantitative way in the following sections.

\begin{figure*}
   \centering
   \includegraphics[width=11cm]{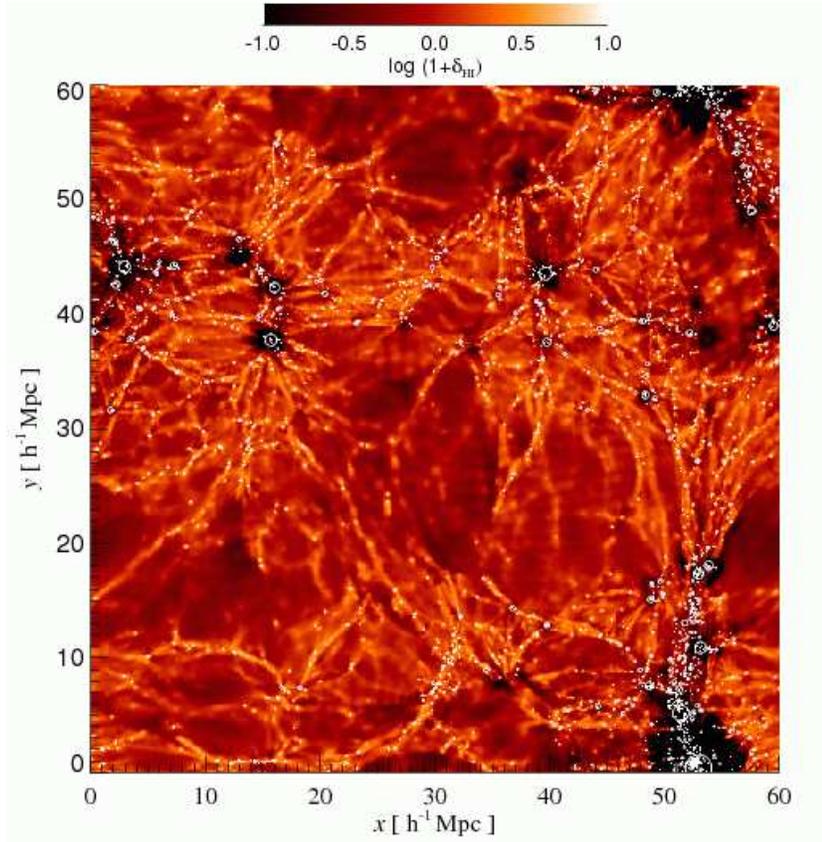}
 \caption{The spatial distribution of the neutral hydrogen overdensity in a slice of
thickness 6 $\hmpc$ (comoving) around the largest cluster in the simulation box (bottom--right part of the panel) extracted from the ESW simulation.
White dots are drawn at the position of dark matter halos. Their size is proportional 
to the halo mass.} 
\label{fig:simul}
\end{figure*}

\subsection{Mock galaxies}

In the simulation we assume a one--to--one correspondence between
gas-rich galaxies and their dark matter halo hosts.  We extract halos
using a friend--of--friend algorithm with a linking length which is 0.2
times the mean interparticle separation and consider only
identified haloes in the mass range $[8 \times 10^{10}, \ 10^{13.5}
\ {\rm M}_{\odot}  \ h^{-1}]$.  The lower limit is set (conservatively) by the numerical
resolution while the upper limit avoids including large halos
associated with groups and clusters, rather than single galaxies. However, we
have checked that including the few halos larger than $10^{13.5}
\ {\rm M}_{\odot}$ does not affect the results presented in this work.  
The geometric mean mass of the haloes is
$\sim 2.46\times 10^{11} \ {\rm M}_{\odot} \ h^{-1}$, to be compared with
a mean mass $10^{11} \ {\rm M}_{\odot} \ h^{-1}$ associated to 
dark matter halos hosting HIPASS galaxies (RW06,  \cite{Mo05}).  
The space density of these mock galaxies (0.0023 per cubic
$\hmpc$ comoving) is  similar to that of 
HIPASS galaxies in the volume limited sample of \cite{Meyer07} [M07]
($\sim 0.003$ per cubic $\hmpc$). This sample contains all galaxies 
within  30 $\hmpc$  and HI mass above
$ 10^{9.05} \ {\rm M}_{\odot} \ h^{-2}$, corresponding to 
a halo mass of $\sim10^{11} \ {\rm M}_{\odot} \ h^{-1}$, as inferred from the 
\cite{Mo05} model, i.e.  similar to our lower mass cut off.
As we shall see in Section~\ref{sec:rspace} the spatial two-point correlation
function of these mock galaxies matches that of the HIPASS objects, hence
fulfilling the main requirement of our analysis.

The mock galaxies extracted from the three simulations
are hosted in the same dark matter haloes that, however, have
a different baryon (gas+star) content.  The baryon mass in the mock 
galaxies is affected by galactic winds and star formation processes.
The mean baryonic mass measured in
the NW, SW and ESW simulations is respectively
$4.7  \, , \  2.1 \  {\rm and} \ 1.9 \times 10^{10} \, \  {\rm M}_{\odot} \ h^{-1}$, 
thus  indicating that galactic winds are quite effective in 
blowing baryons out of  dark halos. The star formation mechanism also plays a role:
the mean stellar mass of  $3.0 \times 10^{10} \, \  {\rm M}_{\odot} \ h^{-1}$ in the NW simulation
decreases to  $0.6 \ {\rm and} \ 0.5 \times 10^{10} \, \  {\rm M}_{\odot} \ h^{-1}$ in the SW and ESW
experiments that adopt the multiphase criterion.
 
To better investigate the dependence of the spatial correlation on the galaxy mass we 
have divided, for the NW case only, the mock galaxy sample by mass in two subsets.
The characteristics of all mock galaxy samples considered
in this paper are summarized in Table~2.

Finally, to compute the correlation properties of the mock galaxies we have generated 
a random galaxy sample by
randomly positioning $5 \times 10^{4}$ objects in the simulation volume.

\begin{table}
\begin{center}
\begin{tabular}{|c|c|c|c|c|c|c|}

\hline 
\multicolumn{7}{|c|}{Mock Galaxy Samples} \\ \hline
Sample & 
N$_{\rm gal} $ & 
${\rm M}_{\rm Min} $ & 
${\rm M}_{\rm Max} $ & 
$ \langle {\rm M}_{\rm DM} \rangle $ &
$ \langle {\rm M}_{\rm bar} \rangle $ & 
Wind  \\ \hline 
G$_{\rm NW}$                    & 4980 & 8.0 & 3160& 24.6 & 4.7 & NW \\ \hline
HG                  & 2480 & 19 & 3160& 53.4 & 10.9& NW \\ \hline
LG                  & 2500 & 8.0 & 19 & 11.4 & 2.0 & NW \\ \hline
G$_{\rm SW}$ & 4980 & 8.6 & 3128 & 25.6 & 2.1 & SW  \\ \hline
G$_{\rm ESW}$ & 4980 & 8.6 & 3100 & 25.4 & 1.9 & ESW \\ \hline
\hline
\end{tabular}
\label{tab:halos}
\caption{
Mock Galaxy Samples. 
Column 1: Sample name. 
Column 2: Number of mock galaxies.
Column 3. Minimum dark  halo mass.
Column 4. Maximum dark halo mass.
Column 5. Geometric mean dark halo mass.
Column 6. Geometric mean baryonic mass.
Column 7. Wind Model.  All masses are in $10^{10} \msun  \ h^{-1}$ units.
}
\end{center}
\end{table}

\subsection{Mock \lya\ absorbers}

The computational box was pierced with 999 straight lines
running parallel to the three Cartesian axes.
Three sets of  333 mock  \lya\ absorption spectra 
along each axis were simulated and analyzed, 
both in real and redshift--space, to measure the 
position of each \lya\ line and the column density of the 
associated HI absorber.
In this work we only consider weak \lya\ absorbers with  column densities in the
range $12.41 \leq \log (\nhi/$\cm$) \leq14.81$ to match the characteristics of the
RW06  sample. 

The total number  of absorbers increases slightly in presence of winds,
while their average column density decreases, as shown in Table~ 3.
However, the differences are small, especially between the SW and ESW 
experiments. The density of \lya\ absorbers  along the line--of--sight 
 in the NW simulation 
($\sim10^{-3} \ {\rm km}^{-1} {\rm s} $) is larger than in the RW06 sample
($\sim 4 \times 10^{-4}   \ {\rm km}^{-1}{\rm s}$). 

To investigate the significance of this mismatch
we have computed the number of
\Lya\ absorbers in our mock spectra, per unit redshift and column
density in each of the three simulations and compared it with that
measured by \cite{Penton04} in the Space Telescope Imaging
Spectrograph (STIS) QSO spectra.  The results are shown in
fig.~\ref{fig:dndz}. The solid, red curve refers to the NW simulation.  The
short--dashed green and the dot-dashed blue curves represent the
\lya\ lines in the SW and ESW runs, respectively. The distribution of
the absorbers is robust to the presence of galactic winds. When
compared to the STIS data of \cite{Penton04} (long--dashed black
curve) we note that the number of absorbers predicted by the
simulation is larger than the observed ones over most of the \nhi\,
range sampled by RW06 (indicated by the two vertical dotted lines).
The difference between models and data, however, is well within
observational errors of $\sim 1$ dex for $ \log (\nhi/$\cm$)
\leq14.5$ (\cite{Penton04}). Since we expect that 
similar observational errors for RW06
absorbers, we conclude that there is no significant difference
in the number density of mock and RW06 \lya\ lines.

To investigate the dependence of the clustering properties
on the absorber column density we have set  a
column density threshold  \nhi\ $ = 10^{13.24}$ \cm
which divides the sample in two equally large subsets 
and sorted all mock absorbers in the NW  simulation
by column density.

The main characteristics of  each mock absorber sample are listed in Table~3.
Moreover, since  these mock sample contains many more spectra than in the real case,
we have also extracted several absorbers' sub--samples of 27 lines--of--sights
to mimic the RW06 sample and assess the sampling noise.

Finally, to compute the two--point spatial correlation functions, we have generated
random absorber samples by randomly positioning 
50 \lya\ absorption lines along  the same 999  lines--of--sight used for the mock \lya\ absorption spectra.
We verified that the estimation of the correlation function does not change significantly 
if, instead, we consider 999 randomly chosen lines--of--sight for the random absorber samples.
We note that 50 lines per spectra represents a good compromise between accuracy and
computing time since doubling the number of random absorbers does not modify 
our estimates of $\xi$. 

\begin{figure}\ 
   \centering
   \includegraphics[width=8.5cm]{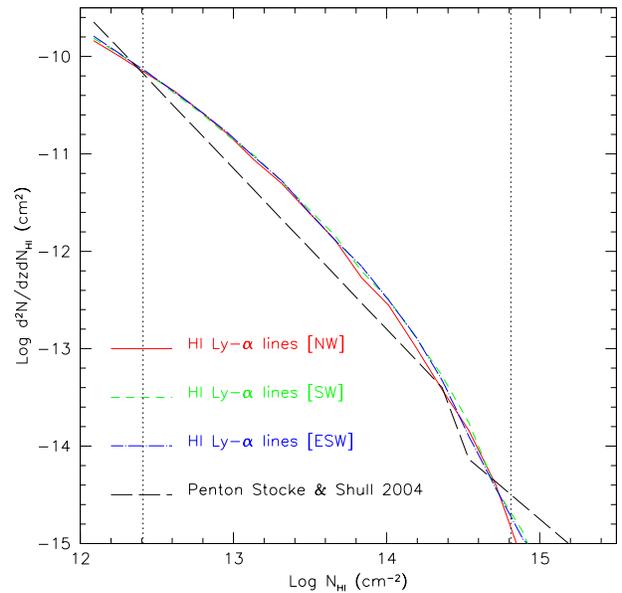}
 \caption{Number of \lya\ absorbers per unit redshift and column density. 
 Solid red line: NW simulation. 
 Dashed green line: SW simulation.
 Dot--dashed blue line: ESW simulation.
 Dashed black:  \citep{Penton04} best fit.}
\label{fig:dndz}
\end{figure}

\begin{table}
\begin{center}
\begin{tabular}{|c|c|c|c|c|c|}
\hline 
\multicolumn{5}{|c|}{Mock Absorber Samples} \\ \hline
Sample & 
N$_{\rm Abs} $ & 
\nhi$_{\rm Min} $ & 
\nhi$_{\rm Max} $ & 
Wind  \\ \hline 
A $_{\rm NW}$         & 6239 & 12.41 & 14.81& NW \\ \hline
HA                  & 1917 & 13.24 & 14.81& NW \\ \hline
LA                  & 4322 & 12.41 & 13.24 &  NW \\ \hline
A$_{\rm SW}$ & 6444 & 12.41 & 14.81 &  SW \\ \hline
A$_{\rm ESW}$ & 6445 & 12.41 & 14.81 & ESW \\ \hline
\hline 
\end{tabular}
\label{tab:lya}
\caption{Absorber Samples. Column 1: Sample name. 
Column 2: Number of mock \lya\ absorbers.
Column 3. Minimum column density.
Column 4. Maximum column density.
Column 5. Wind model.  All column densities are in $\log($\cm$)$ units.
}
\end{center}
\end{table}

\section{Correlation Estimators}
\label{sec:method}

In this work we use 
the \cite{Davis83} estimator to compute the galaxy--absorber
cross--correlation function  both in real and redshift--space,
$\xi(r_p,\pi)$  as:
\begin{equation}
\xi(r_p,\pi)=\frac{AG(r_p,\pi)}{RG(r_p,\pi)}\frac{n_{RG}}{n_{AG}}-1,
\label{eqn:1}
\end{equation}
where $AG(r_p,\pi)$ is the number of mock absorber--galaxy pairs
with projected separation, $r_p$ in the range $[r_p-\delta r_p/2,
r_p+\delta r_p/2]$, and separation along the line--of--sight, $\pi$, in the range
$[\pi-\delta\pi/2,\pi+\delta\pi/2]$.  $RG(r_p,\pi)$ is the
number of pairs consisting of a random absorber and a mock galaxy. 
In both axes the binning $\delta\sigma$ and $\delta\pi$ is set at  0.39 \h\ Mpc,
i.e. 4 times wider than in RW06.
The pair counts are divided by the total number of random--galaxy pairs $n_{RG}$, and
galaxy--absorber pairs, $n_{AG}$. 
The separations $r_p$ and $\pi$ between two objects 
are computed from their recession velocities $\bv_i$  and $\bv_j$
according to \citep{fisher04}:
\begin{equation}
\pi=\frac{\bl \cdot \bs}{H_0|\bl |}, \ \ \ \
r_{p}=\frac{\bs \cdot \bs}{H_0^2}-\pi^2
\label{eqn:reds}
\end{equation}
where $\bl \equiv(\bv_1+\bv_2)/2$ and 
$\bs\equiv \bv_1-\bv_2$.
The estimator~(\ref{eqn:1}) is evaluated in the range of separations [0, 50] \hmpc
both along $r_p$ and $\pi$ directions.
To estimate the galaxy--absorber correlation function in redshift--space
we have used the distant observer approximation, i.e. we have counted
the galaxy--absorber pairs in each of the  three subsets of mock spectra 
parallel to one Cartesian axis and considered only the corresponding 
component of the peculiar velocity to compute the redshift.
The rationale behind this choice is to  detect and average out possible 
geometrical distortions arising,  for example, when 
lines--of--sights are oriented along  HI-rich gas filaments
or when a large fraction of mock galaxies belong to some
prominent, anisotropic cosmic structure.

The galaxy--galaxy and absorber--absorber 
auto--correlation functions are calculated in a similar way, i.e.
by counting galaxy--galaxy and absorber--absorber rather than galaxy-absorbers pairs.
The spherical average of  $\xi(\sigma,\pi)$ gives the spatial correlation function
$\xi(s)$  where $s=\sqrt{r_p^2+\pi^2}$.
We also estimate the analogous quantity in real--space, $\xi(r)$,   where $r$ 
represents the genuine pair separation that coincides wit their
redshift difference  in absence of peculiar velocities.
In order to compare our result with those of RW06 we 
compute two more quantities. The first one is the projected correlation 
function, $\Xi(r_p)$:
\begin{equation}
\Xi(r_p)=2\int_0^{\pi_{max}} \xi(r_p,\pi)d\pi,
\label{eqn:proj}
\end{equation}
where $\pi_{max}=50 \ \hmpc$. 

The second one is the absorber auto--correlation along
individual lines--of--sight,
\begin{equation}
\xi(\pi)=\frac{AA(\pi)}{AR(\pi)}\frac{n_{AR}}{n_{AA}}-1,
\label{eqn:aa}
\end{equation}
where $AA(\pi)$ is the number of mock absorber pairs with separation
$\pi$ along the line--of--sight and $AR(\pi)$ is the number of random absorber pairs. 

The uncertainties in the cross and auto--correlation functions
of the mock samples are computed 
using the bootstrap resampling technique.
For large, independent datasets bootstrap errors are equivalent to  
uncertainties calculated using the jackknife resampling,
as in RW06.
The uncertainty is computed in each $(r_p,\pi)$ bin as
\begin{equation}
\sigma_{\xi_{i}}^2=\frac{\sum_{j=1}^N{\left( \bar{\xi_i}-\xi_i^j\right)}^2 }{N-1}
,\label{eqn:variance}
\end{equation}
where the subscript $_i$ identifies the bin, $_j$ refer the sample and
$\bar{\xi_i}$ is the average correlation function computed over the 
$N$ bootstrapped samples.  In this work $N=50$ which provide us with a 
robust error estimate (increasing $N$ to 350 modifies errors by $<2\%$).

This error estimate assumes that the covariance matrix of the data is 
diagonal, i.e. that the values of $\xi(\sigma,\pi)$  in different bins are not independent, which 
is known not to be the case. However, our simple way of estimating the uncertainties avoids the 
complication of dealing with a large covariance matrix, while providing an unbiased 
estimate of the real errors \citep{Hawkins03}.

\section{Real--Space Analysis}
\label{sec:rspace}

In this analysis we ignore peculiar velocities when we use 
Eq.~\ref{eqn:reds} to estimate $r_p$ and $\pi$ from redshifts.
In Fig.~\ref{fig:xir1} we show the real--space auto--correlation
function of the mock galaxies in the G$_{\rm NW}$ sample (black dots).  Errorbars
represent 1-$\sigma$ bootstrap uncertainties. The autocorrelation of
mock galaxies is shown together with that of HIPASS galaxies,
indicated by the dashed line which represents the power--law best fit to the
$\xi(r)$ in the volume--limited sub-sample of galaxies 
extracted from the HIPASS catalog by M07.
This power--law has a  slope $\gamma_{gg}=1.5\pm1$ and correlation length $r_{0,gg} = 3.2 \pm 1.4$ \hmpc.
The two functions agree, within the errors, below $10 \ \hmpc$, since
the  power--law fit to the correlation function of our  mock galaxies  in the range  $[1, 10] \ \hmpc$
has  $\gamma_{gg}=1.46\pm 0.03$ and  $r_{0,gg} = 3.06 \pm 0.15$ \hmpc.
We have considered the M07 result since it is based on a sub-catalog that
is volume limited, like our mock samples but it is worth noticing that
the RW06 fit obtained using the full, flux limited HIPASS sample
is fully consistent with the M07 result and, therefore, with our 
fit too.

\begin{figure}\ 
   \centering
   \includegraphics[width=8cm,angle=-90]{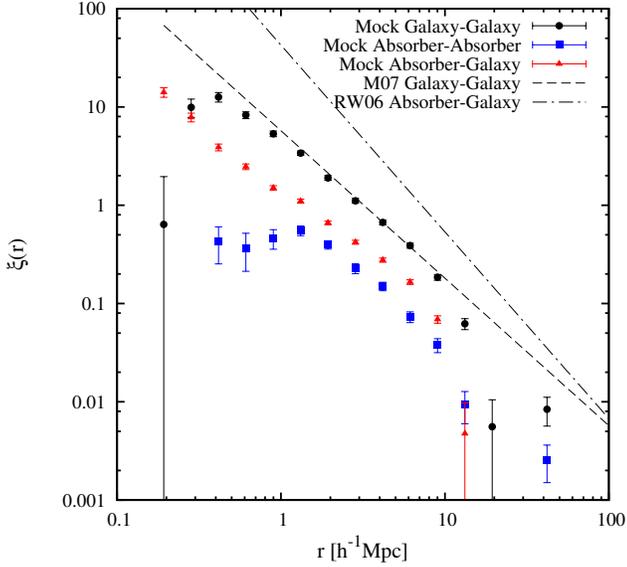}
\caption{
The real--space two--point correlation functions
of the mock galaxy and absorber samples. 
Black dots: galaxy auto--correlation function in the 
G$_{\rm NW}$-sample. Blue squares: absorber autocorrelation function 
in the A$_{\rm NW}$--sample. Red triangles: galaxy--absorber autocorrelation
function in the A$_{\rm NW}$+G$_{\rm NW}$ sample. The size of the bars shows 1-$\sigma$ 
bootstrap resampling uncertainties. 
Black dashed curve: best fit to the galaxy--galaxy 
correlation function in the HIPASS volume--limited sample of M07.
Black  dot--dashed curve: RW06 best fit to the HIPASS galaxy- \lya\
absorbers cross--correlation function.
}
\label{fig:xir1}
\end{figure}

The correlation signal of the mock galaxies suddenly drops 
at separations smaller than 0.4 \hmpc.
On the contrary,  the galaxy correlation function of RW06
monotonically increases when reducing the pair  separation.
Including the few mock halos larger than $10^{13.5} \msun$ 
sample does not modify significantly this small-scale trend.
This small--scale mismatch as an artifact deriving 
from the fact that, in the simulation, we do not 
resolve galaxy--size sub--structures 
within the large cluster--size halos that, if present, would significantly 
contribute to the correlation signal at sub--\hmpc scales.
Indeed, when we run the Friends--of--Friends algorithm 
to identify halos using a smaller linking length of 0.1 times the mean 
inter--particle spacing, the small scale flattening disappears and the 
power--law behavior is restored below 0.3 \hmpc.

Mock absorbers are significantly less self--clustered than galaxies:
their autocorrelation function (blue squares) is factor of $\sim 10$
below that of galaxies (see \cite{Dobrzycki02}).  We cannot compare this result with
observational data directly, since the observed \lya\ absorbers are too
sparse. However, RW06 was able to compute their correlation along each
line--of--sight and we compare this result with the theoretical predictions in
the next Section.

The red triangles show the mock galaxy--absorber cross--correlation function
of the A$_{\rm NW}$+G$_{\rm NW}$ samples which 
is  significantly  weaker than the galaxy auto--correlation.
This result is at variance with that of RW06 who find that
the  cross--correlation function of HIPASS galaxies and \lya\ absorbers
(dot--dashed curve in Figure~\ref{fig:xir1}) 
in the  $[1,10]$ \hmpc range is best fitted with a power--law 
slope  $\gamma_{ag}=1.9\pm 0.3$ and correlation length $r_{0,ag} = 7.2 \pm 1.4$ \hmpc,
significantly  larger than that of the galaxy auto--correlation function.
When we fit  the cross--correlation function of the mock data in the same 
range of separations we find
 $\gamma_{ag}=1.29\pm 0.03$ and correlation length $r_{0,ag} = 1.44 \pm 0.08$ \hmpc.

RW06 pointed out that the cross--correlation signal increases with the
column density of the absorber.  We find the same trend in the
simulation. We show in Fig.~\ref{fig:xr2}, the cross--correlation
signal increases when we restrict  our analysis to strong absorbers of the HA
sample.  On the contrary, massive mock galaxies do not seem to be
significantly more or less correlated to \lya\ absorbers than smaller galaxies. In fact,
we find that the cross--correlation signal is almost independent of galaxy mass.

\begin{figure}\ 
   \centering
   \includegraphics[width=8cm,angle=-90]{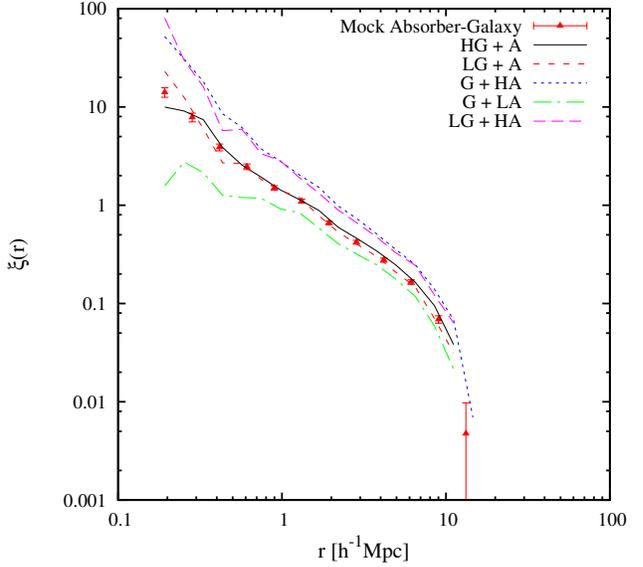}
\caption{The absorber--galaxy
 cross--correlation function in various mock subsamples.
Red triangles with errorbars:  A$_{\rm NW}$+G$_{\rm NW}$ sample. Black solid line:
HG+A$_{\rm NW}$.
 Dashed red: LG+A$_{\rm NW}$. Dotted blue: G$_{\rm NW}$+HA. Dot-dashed green:
G$_{\rm NW}$+LA. Long-dashed magenta: LG+HA.}
\label{fig:xr2}
\end{figure}

Strong galactic winds can blow gas out of galaxy--size halos and
therefore could suppress the cross--correlation signal on sub-Mpc
scales.  To quantify the effect we have computed the galaxy--absorber
correlation functions in the SW and ESW simulations and compared them with
that of the NW experiment. The results are shown in
Fig.~\ref{fig:xr3}. The red triangles with errorbars represent the
same cross--correlation function of the A$_{\rm NW}$+G$_{\rm NW}$ sample shown in
Fig.~\ref{fig:xr2} and refer to the case of no winds.  The effect of
including the effect of strong winds is illustrated by the blue dashed
and solid black curves that refer to the SW and ESW simulations,
respectively.  Even adopting extreme prescriptions for galactic winds,
the effect on the galaxy--absorber correlation function is very small
and, as expected, is significant only at separations $\lesssim 0.3\,
\hmpc$ where fewer galaxy--absorber pairs are found with respect to the
NW case. This is not surprising, considering the   free travel length
$l_w$ adopted in the models.
We find no significant differences between the SW and ESW
experiments, which illustrates the robustness of the cross--correlation
signal  on scales larger than $l_w$
to the scheme adopted to simulate galactic winds.

\begin{figure}\ 
   \centering
   \includegraphics[width=8cm,angle=-90]{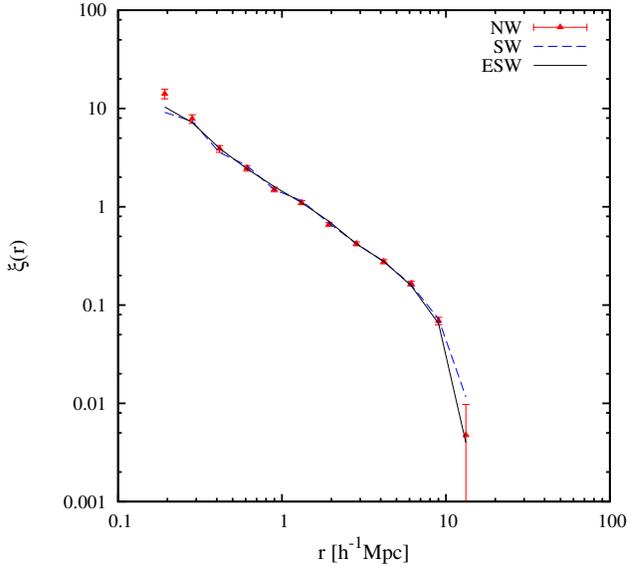}
\caption{Effect of galactic winds on the 
cross-correlation function. 
Red triangles: A$_{\rm NW}$+G$_{\rm NW}$ samples in the NW simulation.
Dashed blue curve: SW simulation.
Solid, black curve: ESW simulation.
The size of the bars shows 1-$\sigma$ 
bootstrap resampling errors.  
}
\label{fig:xr3}
\end{figure}

\section{Redshift--Space Analysis}
\label{sec:zspace}

In Section~\ref{sec:rspace} we have shown that hydrodynamical
simulations do not reproduce the RW06 result. On the contrary, the
galaxy-absorber correlation function is significantly weaker than the
galaxy autocorrelation function.  The previous analysis, however, has
been performed in real--space ignoring peculiar velocities that may
bias the correlation analysis.
Moreover we have considered a number of  spectra  much larger than
that of RW06.  
Therefore, we must account for the
 possibility is that the mismatch between hydrodynamical simulations
and RW06 is not genuine but derives, instead, from redshift--space distortions
and sparse HI sampling that, if not properly accounted for, may affect
the cross-correlation analysis.  In an attempt to account for both types of 
errors we repeat the correlation analysis using more realistic mock
catalogs in which redshifts are used as distance indicators and only 
 27 lines-of-sight are taken to mimic the RW06 data set.
To investigate the two effects separately, we first
perform a redshift--space analysis of the whole A$_{\rm NW}$+G$_{\rm NW}$
 sample and then we consider  sub-samples of 27 lines-of-sight.

In Figure \ref{fig:galgal2D} the autocorrelation function of the mock galaxies in the 
G$_{\rm NW}$ sample, 
$\xi_{gg}(r_p,\pi)$, is 
plotted on the $(r_p,\pi)$ plane. Contours are drawn at iso--correlation levels
of 2,1,0.5,0.25. The distortions along the $\pi$ axis induced by 
small scale incoherent motions within virialized structures (the so called fingers--of--god) 
can be seen at separations $r_p \le 2\, \hmpc $  extending out to $\pi = 6$ $\hmpc$.
A similar distortion pattern is seen in the correlation function of HIPASS
galaxies (Fig. 2 of RW06). In that case a second, independent, distortion pattern  
along  the $r_p$ axis, is detected at separations 
$r_p \gtrsim 4\, \hmpc$.  The compression of the isodensity contours along $\pi$ 
is the signature of large scale coherent motions that  increase the apparent number of pairs 
with large separations. This second distortion pattern is not visible in
Fig.~\ref{fig:galgal2D}, a fact that we ascribe to the lack of large scale power in
our simulations. Indeed, our simulations do not account for power on scales
larger than 60 \hmpc which could significantly contribute to the amplitude of the 
bulk motions and thus to the compression of the iso--density contours.

\begin{figure}\ 
   \centering
   \includegraphics[width=12cm,angle=-90]{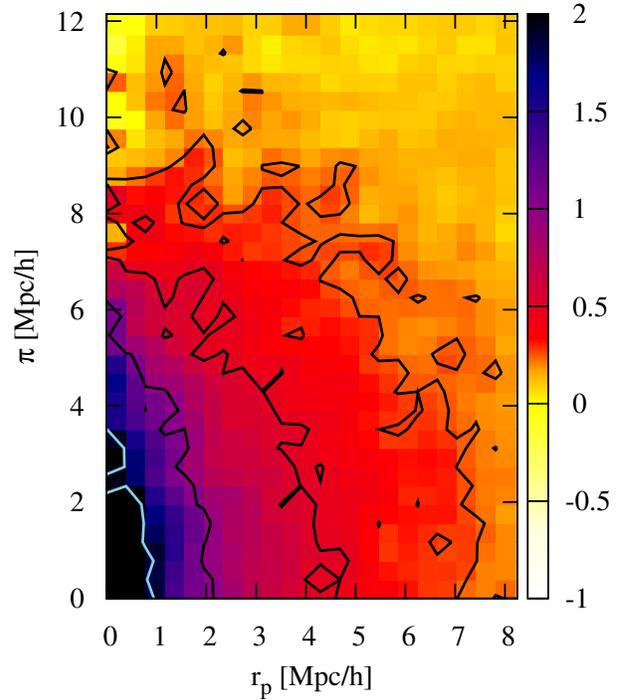}
\caption{The redshift--space auto--correlation function
$\xi_{gg}(r_p,\pi)$ of mock galaxies in the G$_{\rm NW}$ sample.
The binning is 0.4 \hmpc in both axis. 
Iso-correlation contours are drawn at correlation levels
of 2,1,0.5,0.25.
Bootstrap resampling shows typical pixel variations of the order 
 $\xi(r_p,\pi) \sim 0.08 $}
\label{fig:galgal2D}
\end{figure}

Figure~\ref{fig:galabs2D} shows the redshift--space cross--correlation function
$\xi_{ag}(r_p,\pi)$ of mock absorbers and galaxies in the A$_{\rm NW}$+G$_{\rm NW}$ sample.
The signal is significantly weaker than the galaxy auto--correlation and 
the distortion pattern looks very different as  no significant 
elongation is seen along the $\pi$ axis.  Instead, at large separations, 
the iso--correlation contours are  compressed along  $\pi$, as expected in the
presence of  coherent motions.
The differences between $\xi_{ag}(r_p,\pi)$ and $\xi_{gg}(r_p,\pi)$
reveal that mock  \lya\ absorbers  and galaxies have different dynamical properties.
Galaxies' relative velocities are  dominated by the incoherent motions, typical 
of virialized structures.
Instead, the relative motion of mock \lya\ absorbers and galaxies is more
coherent, suggesting that mock absorbers are preferentially located in the outskirts
of high density regions into which they are probably falling.

Finally, we note that the peak of the cross-correlation function  is spatially offset 
from the center. This feature and the general distortion pattern of the
simulated cross-correlation function are qualitatively
 similar to that of the cross-correlation function 
between the CHFT galaxies and the  $HST$ Quasar Absorption Line Key Project 
Data Release \lya\  with $13\le \log(\nhi/$\cm$) < 15$ measured by W07.
On the contrary, the RW06 cross-correlation function is dominated by a
 very large finger--of--god distortion. A similar, but less prominent, distortion pattern
has been seen by  \cite{Dave99} and W07 in their numerical experiments.
 RW06 interpreted this distortion  as the draining 
of the gas from low-density regions into collapsed structure. Although the dynamical
interpretation in this case is not as simple as  in the galaxy-galaxy case, we note 
that the draining mechanism advocated by RW06 would probably lead to coherent,
rather than incoherent motions, which would produce a very different distortion pattern.
W07 suggested that the finger--of--god distortion  could  be a geometrical effect deriving
from observing \lya\ absorbers along  lines--of--sights that run along some radially-elongated structure.
To check this hypothesis we exploited the distant observer approximations and 
computed the cross-correlation function by considering redshift distortions along one Cartesian 
axis at a time. 
If  distortions were purely geometric, i.e. induced 
by a few prominent, anisotropic structures,  
we would  expect to see different distortion patterns
in the cross-correlation functions computed along 
orthogonal  axes. If, on the other hand, they were caused by 
random motions within large, spherically symmetric, virialized structures like 
galaxy clusters, we would expect to see fingers-of-god type distortions along 
all axes. Instead, the correlation functions measured by three, orthogonally--positioned
distant observers turned out to be very similar and consistent with the one 
shown in Fig.~\ref{fig:galabs2D}.
We conclude that neither pure geometrical effects nor incoherent motions
can alone explain the distortion pattern in the $\xi_{ag}(r_p,\pi)$ of our mock A$_{\rm NW}$+G$_{\rm NW}$ samples.

\begin{figure}\ 
   \centering
   \includegraphics[width=12cm,angle=-90]{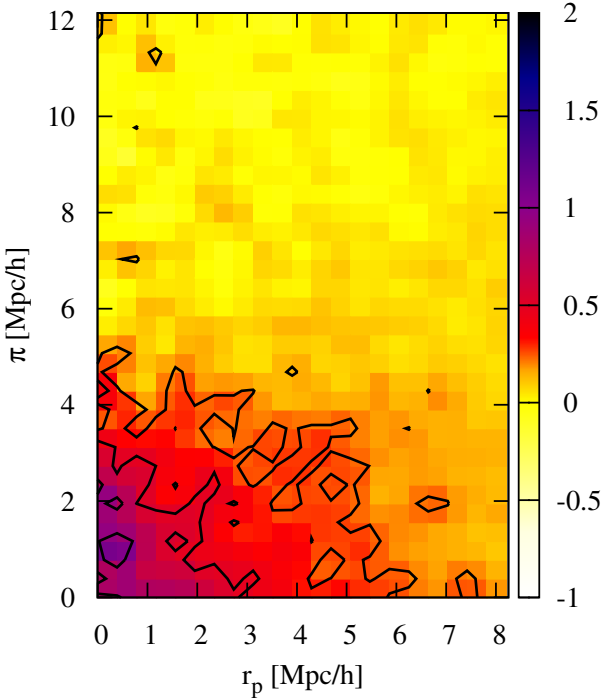}
\caption{The redshift--space cross--correlation function
$\xi_{ag}(r_p,\pi)$ of mock absorbers and galaxies in the A$_{\rm NW}$+G$_{\rm NW}$ sample.
The binning is 0.4 \hmpc in both axis. 
Iso-correlation contours are drawn at correlation levels
of 1,0.5,0.25.
Bootstrap resampling shows typical pixel variations of the order 
 $\xi(r_p,\pi) \sim 0.2 $}
\label{fig:galabs2D}
\end{figure}

Small redshift distortions   could be amplified 
by sampling \lya\ absorbers along  a limited number of lines--of--sights, as in the RW06 case.
To quantify the effect of  shot noise errors coupled to dynamical and geometrically--induced
distortions, we have constructed 30 independent realistic mock \lya\ sub-samples 
of 27 independent lines--of--sights and computed their
cross--correlation with all mock galaxies of the G$_{\rm NW}$ sample.
In Fig.~\ref{fig:4cross} we show $\xi_{ag}(r_p,\pi)$  computed in four such
realistic mock samples. The cross-correlation functions shown in the two upper panels
are characterized by  prominent finger--of--god distortions which, in the upper--right plot,
are similar in amplitude to that measured by RW06.
This kind of distortion is found in $\sim 20 \%$  of the mock subsamples considered.
The fact that we observe fingers--of--god distortions  along different Cartesian axes
suggest that they cannot be attributed to the fact that the sample is dominated by a 
single, prominent, anisotropic structure. Rather, they seem to originate from 
genuine, finger--of--god like, dynamical distortions which become apparent when 
a significant fraction of the 27 spectra samples some virialized regions.
The relevance of sparse  sampling variance in the cross-correlation analysis is even more 
evident in the two bottom panels of Fig.~\ref{fig:4cross}.  They show the cross--correlation 
function computed along the same ($Z-$) axis, as in  
the top--right panel, but use two independent sets of lines--of--sight.  
Not only  the finger--of--god distortion disappears but the cross--correlation signal is either 
very weak (bottom left) or significantly offset from the center (bottom-right). 

A more quantitative assessment of sparse sampling errors is given in
Fig.~\ref{fig:proj} in which we show the projected absorber-galaxy
cross correlation function $\Xi_{ag}(r_p)/r_p$ of the A$_{\rm NW}$+G$_{\rm NW}$ sample
(filled black dots).  Small errorbars drawn with solid lines represent
1-$\sigma$ bootstrap resampling errors computed using all 999 mock
absorbers in the A catalog.  Large errorbars plotted with dashed lines
represent the scatter around the mean of the projected
cross--correlation function computed using the 30 realistic mock
absorbers' samples consisting of 27 lines--of--sight.  The sampling
noise clearly dominates the error budget and the total error
significantly exceeds that of RW06.  Filled red squares
show the projected galaxy-galaxy correlation function with the
1-$\sigma$ bootstrap errors. 
In order to assess the goodness of our error estimate we have
compared the scatter among the 30 catalogs with the bootstrap errors
computed from N=50 samples. The two errors agree well in the range
(1,10) Mpc/h, in which boostrap errors are $\sim 15 \ \%$ smaller than those shown
in Fig. ~\ref{fig:proj}. On smaller scales the bootstrap resampling technique
overestimates the errors by factor of $\sim 2$.
The autocorrelation signal is higher than
the cross--correlation one, consistently with the real--space
analysis.  However, the difference is of the order of the errors, i.e.
the mismatch is about 1-$\sigma$ at separations $r_p>1\,\hmpc$,
 in the range in which RW06 find that the
cross-correlation signal is larger than the autocorrelation one.
Filled triangles show the projected autocorrelation function of all
absorbers in the A$_{\rm NW}$ sample.  As anticipated by the real--space
analysis, absorbers correlate with themselves very weakly.  When one
accounts for sparse sampling their autocorrelation signal is
consistent with zero.

RW06 were able to detect the auto-correlation signal of the absorbers by measuring 
their auto-correlation function $\xi(\pi)$ of eq.~\ref{eqn:aa} along individual lines--of--sight.
We have repeated that analysis using all absorbers in the A sample. 
{The resulting auto-correlation function replicates the RW06 result to within 1-$\sigma$}.

Finally, to test the robustness of our results we have  computed  the cross--correlation
function, $ \xi_{ag}(r_p,\pi)$ using the HG, LG, and LA sub--samples as well as
the mock catalogs extracted from the SW and ESW runs. There are no cases in which wer are able to
obtain a galaxy--galaxy autocorrelation signal weaker than the cross--correlation 
one and to reproduce the large finger--of--god distortion feature observed by RW06.

\begin{figure*}\ 
   \centering
   \includegraphics[width=16cm,angle=-90]{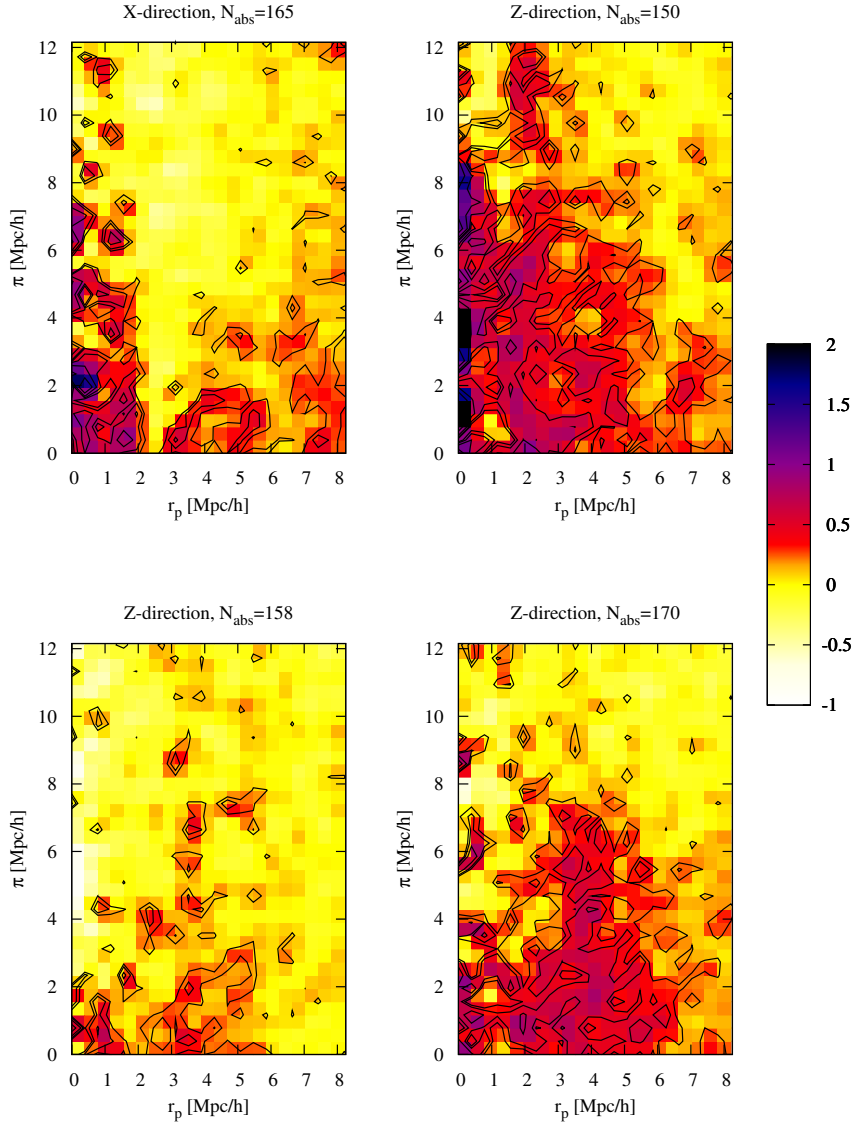}
\caption{The redshift--space
absorber--galaxy cross--correlation function
$\xi_{ag}(r_p,\pi)$ for four independent subset of 
absorbers along 27 lines--of--sight. On top each panel is 
indicated the direction along which absorptions spectra were drawn
and the total number of mock absorbers in each sample N$_{\rm abs}$.
All cross--correlation functions are  computed using the same 4980 mock 
galaxies in the G sample.}
\label{fig:4cross}
\end{figure*}

\begin{figure}\ 
   \centering
   \includegraphics[width=8cm,angle=-90]{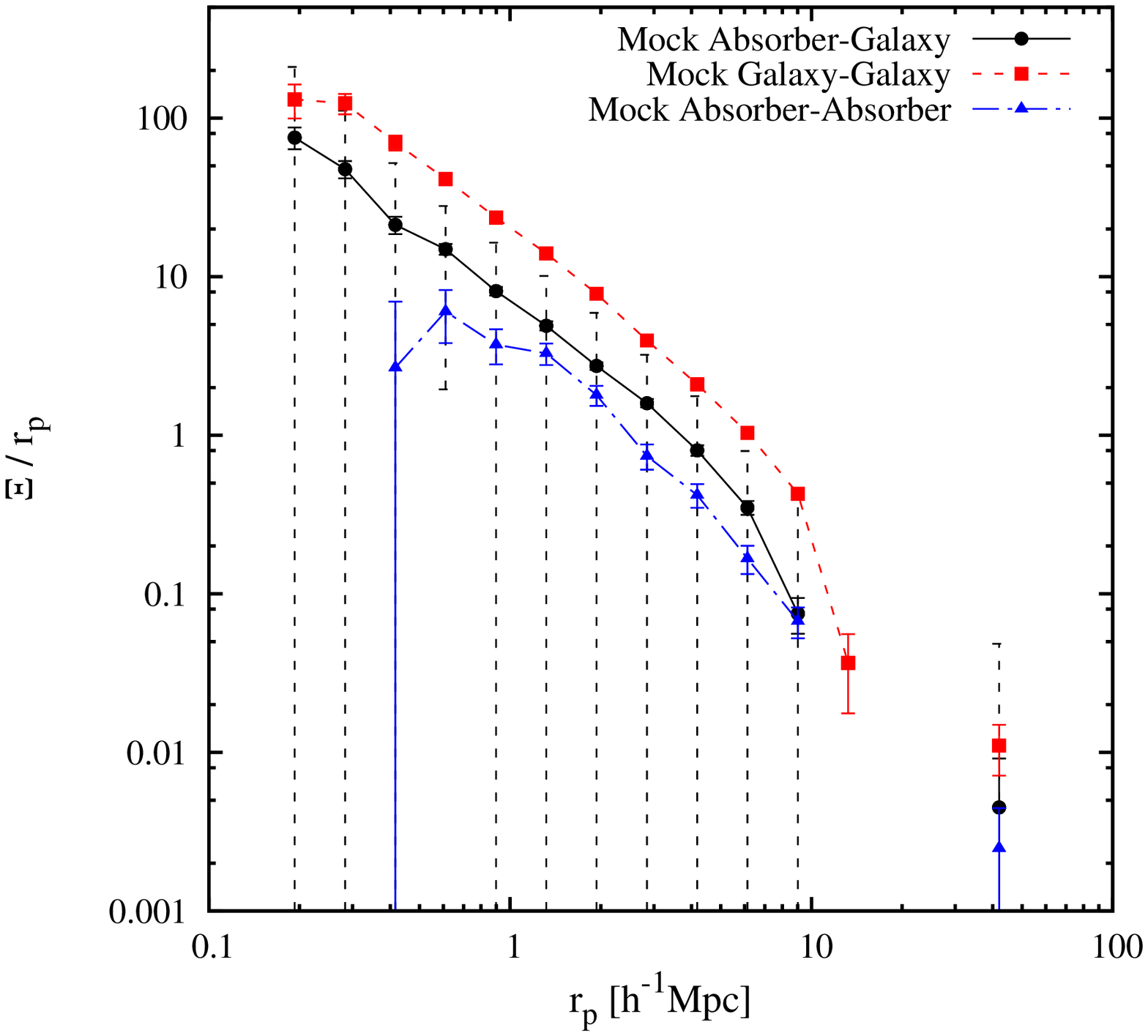}
\caption{Filled dots and solid line:  projected absorber-galaxy cross 
correlation function
$\Xi_{ag}(r_p)/r_p$. Small errorbars are
1-$\sigma$ bootstrap resampling errors. Large errorbars account for
sparse sampling variance. Filled squares and dashed line: galaxy-galaxy projected 
autocorrelation function $\Xi_{ag}(r_p)/r_p$. Filled triangles and 
dot-dashed line: absorber-absorber projected 
autocorrelation function $\Xi_{aa}(r_p)/r_p$}
\label{fig:proj}
\end{figure}

\section{Conclusions}
\label{sec:dis}

In this work we have studied the relative spatial distribution of galaxies and
weak \lya\ absorbers with  $12.41 \leq \log (\nhi/$\cm$) \leq14.81$  
in hydrodynamical simulations and compared our results
with the analyses of real datasets performed by W07 and, mainly, with RW06.
Our main conclusions are:

\begin{itemize}

\item The galaxy-absorber two-point cross-correlation function in the
  hydrodynamical simulation is weaker than the galaxy auto-correlation
  function. This result is at variance with that of RW06 but in
  qualitative agreement with the analysis of W07.

\item  No flattening at small separation is observed in the
  cross correlation function of all mock absorbers, unlike in RW06. 
  A small-scale flattening is observed, however, when the cross correlation
  analysis is restricted to low density absorbers.
 
\item The cross correlation signal increases with the column density
  of the absorbers, in agreement with RW06. We find no significant
  dependence on galaxy mass.

\item Galactic winds have a small effect on the absorber and galaxies
  correlation properties in these models.  Using the most extreme
  prescription to simulate these winds suppresses the
  cross-correlation signal only at separations $\lesssim 0.3 \ \hmpc$.

\item Absorbers correlate with themselves more weakly than with
  galaxies.  Their auto--correlation signal is very weak and
  consistent with that measured by RW06.

\item Redshift--space distortions alone cannot explain two aspects of
  the differences with the RW06 results.  The cross--correlation
  signal is weaker than the galaxy auto--correlation signal.  The
  two-point cross--correlation function, $\xi_{ag}(r_p,\pi)$ does not
  show a prominent finger--of--god type of distortion. The latter
  looks very prominent in the RW06 cross--correlation function but is
  not seen in the W07 one.

\item The origin of the finger--of--god distortion cannot be purely
  geometric, i.e induced by the presence of a prominent, anisotropic
  structure in the sample.  In this case distant observers taking
  spectra along orthogonal directions would detect different
  distortion patterns. We do not see such effect.
 
\item Fingers--of--god distortions may appear when sampling the
  intergalactic gas using a limited number of UV spectra, as in the
  RW06 sample. In this case, they represent genuine dynamical
  distortions that become apparent when a few spectra, that however
  represent a significant fraction of the total, pierce some
  virialized regions.

\item The sampling noise is large. Once accounted for, the difference
  between the simulated galaxy--galaxy and galaxy--absorber
  correlation functions is significant at the $\sim 1$-$\sigma$ level only.

\end{itemize}

\section{Discussion and perspectives}
\label{sec:per}

Modeling the gas distribution in the low redshift universe is a
difficult task.  Numerical experiments use a number of simplifying
hypothesis and approximations that potentially affect our results.
The main uncertainties are related to the ill-known mechanisms of
stellar feedback and galactic winds for which we have adopted
simplistic phenomenological prescriptions.  It is therefore very
reassuring that our results are robust to the star formation criterion
and galactic wind prescriptions adopted.  However, since robustness
does not exclude systematic errors 
one needs to be aware that the
various approximations adopted in our numerical model to predict the
HI distribution at $z=0$ may bias our results.

Our model does not include halos larger that $10^{13.5} \ \msun$ and
ignores substructures within virialized halos. While we have checked
that including large halos does not change our results, ignoring
galaxy-size halos within groups or clusters may affect the outcome of
the correlation analysis.  Galaxies in strongly clustered environments
significantly contribute to both the auto- and the cross-correlation
function at small separations. Ignoring their presence would
artificially decrease the correlation signal, producing a flattening
in the correlation functions at small separations.  We do see a
flattening but only in the galaxy-autocorrelation function and on
scales smaller than 0.4 \hmpc.  The cross--correlation function,
instead, increases at small separations unlike the one of RW06 that
flattens and we do not reproduce the flattening at separations smaller
than 1 \hmpc.  A flattening of the galaxy-absorber cross-correlation
function at small scales was also seen in the numerical simulations of
\cite{Dave99} that, however, have a limited resolution compared to
ours.  The fact that we find no flattening in the cross-correlation
function has two implications. First, ignoring sub-clustering within
large halos has little impact on our results.  Second, it seems that
there is no characteristic scale for the cosmic structures in which
\lya\ absorbers are embedded.

The RW06 analysis convincingly rules out minihaloes for the
confinement of weak \lya\ absorbers. Based on the measured
cross-correlation strength, RW06 suggest that they are embedded in
much larger halos with the typical mass of a galaxy-group.  This would
imply a self--clustering of the absorbers comparable or even larger
than that of galaxies.  The fact that, on the contrary, the measured
absorber self--clustering along the line--of--sight is weak is not
regarded by RW06 as a conclusive evidence since redshift distortions
may artificially dilute the correlation signal.  Our numerical
experiments provide a direct estimate for the self--clustering of the
absorbers which is free of redshift--distortions. The real--space
analysis we have performed indicates that the auto-correlation
function of the mock absorbers is significantly weaker than that of
mock galaxies and that, in redshift--space, their self--clustering is
consistent with the RW06 estimates.  The outcome of our numerical
model suggest therefore that in a $\Lambda$CDM universe weak
\lya\ absorbers are not embedded in group-size halos. In fact, the
association of weak \lya\ absorbers with virialized halos is probably
too naive. The absence of a strong finger--of--god distortions in the
simulated $\xi_{ag}(r_p,\pi)$ suggest that the neutral hydrogen
responsible for weak \lya\ absorption lines is not part of virialized
structures. Rather, it is probably located in their outskirts,
in--falling towards their central regions.  Interestingly, we see a
flattening in the absorber auto--correlation function at separations
$\lesssim 1\, \hmpc$ a feature which is also typical of the the
warm--hot intergalactic gas according to both numerical \citep{Dave01}
and semi-analytic \citep{Valageas02} predictions.

Finally, we turn to what we regard as the main result of this
work. RW06 find that the galaxy-absorber cross-correlation signal is
significantly larger than the galaxy-galaxy correlation. Our numerical
analysis is not able to reproduce the observation as we find that the opposite is true. 
However, when shot
noise errors are accounted for, the discrepancy between the auto- and cross-correlation
signals is of the order of 1-$\sigma$ only. 
Can we reconcile the two results ? 
Our numerical experiments were performed on a rather small box of
$60 \ \hmpc$ which cannot be regarded as a fair sample of the
universe. In other words our cosmic variance is not negligible and
should be accounted for in our error budget. This would require
running numerical simulations in a larger box while keeping the same
resolution or running several identical simulations of different
random realizations of the universe.  In either case the likely
outcome would be that of increasing the size of the errorbars in
fig.~\ref{fig:proj} and the conclusion would be that, probing the HI
distribution with 27 lines--of--sight is not sufficient, in a
$\Lambda$CDM, universe to demonstrate a difference between the self
and cross clustering of galaxy and \lya\ absorbers at the level
measured by RW06.

The fact that the errorbars in the projected cross-correlation
function of RW06 are smaller than ours seem to indicate that
their error estimates are biased low. In section~\ref{sec:zspace}
we have shown that the bootstrap technique underestimates
errors by $\sim 15$ \%, on average, at separations  $ \ge 1 \hmpc $
when the sampling is as sparse as in the RW06 case.
This bias reflects the fact that absorbers are not guaranteed to 
be independent.
It is plausible that this effect is even more severe in the RW06 sample 
since nearly 30\% of  the absorption spectra considered  
were drawn in the vicinity of the Virgo cluster region.
We would also expect that these spectra could 
artificially amplify the cross--correlation signal
since the Virgo cluster is an HI-rich region. 
However, surprisingly enough, the excess
cross-correlation signal is still present when galaxies and absorbers
from this region are excluded from the analysis (Ryan-Weber, private
communication).

The only way out at this apparent paradox is that the relative
distribution of galaxies and \lya\ absorbers in the RW06 sample is
different from that of the typical cosmic environment, since the
cross-correlation signal and its variance are significantly different
from their average values. This despite the fact that in our cosmic
neighborhood the most prominent structures are anisotropically located
along the Super-Galactic plane, rather than being homogeneously
distributed.  We see two possible ways to check the validity of this
hypothesis.  One is to resort to the so called constrained
hydrodynamical experiments  designed to match the actual gas
distribution in our local universe \citep{Kravtsov02, Klypin03,
  Yoshikawa05, Viel05}. Currently available simulations, however, are
of little use as their constraints are either too weak, as they refer
to scales larger than 5 \hmpc (Gaussian), or too local, as they are
effective out to distances of $\sim 15$ \hmpc, i.e. within our local
Supercluster.  
The second possibility, which looks more promising, is
to improve the sampling of the HI distribution either through
\lya\ absorption lines in the UV absorption spectra or through the
X-ray lines of highly ionized metals, like OVII. The latter
is expected to trace the Warm Hot Intergalactic Medium (WHIM)
in density-temperature environment similar to that in which
the weak \lya\ absorbers can be found. With this respect, proposed X-ray
satellites like EDGE \citep{Piro07} are particularly interesting, as
they could observe the WHIM in emission, which would allow one to
trace the three dimensional gas distribution rather than probing it in
1D along a few lines--of--sight.  

It is worth stressing that the present tension between model and data
could be a signature of the fact that hydrodynamical simulations are still
missing physical inputs able to reproduce the observations.
However, if the mismatch between theory and observations is confirmed, which probably requires
both better observational data and better control over systematics 
in the numerical models, the RW06 results could
constitute an interesting challenge to the $\Lambda$CDM paradigm,
similar, and perhaps related to the absence of dwarf galaxies in voids
\citep{Peebles07}.

\section*{Acknowledgments}

The authors thank Simon White for the useful comments on a
version of this manuscript and Emma Ryan-Weber for the
fruitful discussions and suggestions. EB thanks the Max Planck Institute f\"ur
Astrophysik for hospitality when part of this work was done.
Numerical computations were done on the COSMOS supercomputer at DAMTP
and at High Performance Computer Cluster (HPCF) in Cambridge (UK).
COSMOS is a UK-CCC facility which is supported by HEFCE, PPARC and
Silicon Graphics/Cray Research.

\bibliographystyle{mn2e}
\bibliography{mn-jour,paper}

\bsp

\label{lastpage}

\end{document}